\documentclass[prd,amsmath,showpacs,preprintnumbers,superscriptaddress,
floatfix,twocolumn]{revtex4}
\usepackage{amsmath}
\usepackage{graphicx}

\newcommand{\adj}{\dagger}
\newcommand{\oa}[1]{${\cal O}(a^{#1})$}

\newcommand{\tr}{\text{Tr}}

\newcommand{\gambar}{\overline{\gamma}}
\newcommand{\deltabar}{\overline{\delta}}
\newcommand{\chibar}{\overline{\chi}}

\newcommand{\eref}[1]{Eq.~(\ref{#1})}

\begin{document}

\preprint{\vbox{
\rightline{ADP-04-23/T606}
\rightline{}}}

\title{Unquenched quark propagator in Landau gauge}
\author{Patrick O.~Bowman}
\affiliation{Nuclear Theory Center, Indiana University, Bloomington IN 47405, 
USA}
\author{Urs M.~Heller}
\affiliation{American Physical Society, One Research Road, Box 9000,
Ridge NY 11961-9000, USA}
\author{Derek B.~Leinweber}
\author{Maria B.~Parappilly}
\author{Anthony G.~Williams}
\author{Jianbo~Zhang}
\affiliation{Special Research Centre for the Subatomic Structure of Matter 
and \\ Department of Physics, University of Adelaide, Adelaide, SA 5005, 
Australia}

\date{\today}

\begin{abstract} 
We present an unquenched calculation of the quark propagator in Landau gauge 
with 2+1 flavors of dynamical quarks.
We use configurations generated with an improved staggered (``Asqtad'') action 
by the MILC collaboration.  This quark action has been seen to have excellent
rotational symmetry and scaling properties in the quenched quark propagator.
Quenched and dynamical calculations are performed on a $20^3\times 64$ lattice 
with a nominal lattice spacing of $a  = 0.125$ fm.  The matched quenched and 
dynamical 
lattices allow us to investigate the relatively subtle sea quark effects, and 
even in the quenched case the physical volume of these lattices gives access 
to lower momenta than our previous study.  We calculate the quark mass 
function and renormalization function for a variety of valence and sea quark 
masses. 
\end{abstract}


\maketitle

\section{Introduction}
Quantum Chromodynamics is widely accepted as the correct theory of the strong
interactions and the quark propagator is its most basic quantity.
In the low momentum region it exhibits dynamical chiral symmetry breaking 
(which cannot be derived from perturbation theory) and at high momentum it can be 
used to extract the running quark mass~\cite{Bowman:2004xi}. 
In lattice QCD, quark propagators are 
tied together to calculate hadron masses and other properties.  Lattice gauge 
theory provides a way to calculate the quark propagator directly, providing 
access to quantities such as operator product expansion (OPE) condensates~\cite{Arriola:2004en}. In 
turn, such a calculation can provide technical insight into lattice gauge 
theory simulations.

The systematic study of the quark propagator on the lattice has also
provided fruitful interaction with other approaches to hadron physics,
such as instanton phenomenology~\cite{Diakonov:2002fq}, chiral quark
models~\cite{Arriola:2003bs} and Dyson-Schwinger equation
studies~\cite{Bhagwat:2003vw,Alkofer:2003jj}.  The lattice is a first
principles approach and has provided valuable constraints for model
builders.  In turn, such alternative methods can provide feedback on
regions that are difficult to access directly on the lattice, such as
the deep infrared and chiral limits.

The quark propagator has previously been studied using 
Clover~\cite{Skullerud:2000un,Skullerud:2001aw}, 
staggered~\cite{Bowman:2002bm,Bowman:2002kn} and 
Overlap~\cite{Bonnet:2002ih,Zhang:2003fa} actions.  For a review, 
see Ref.~\cite{Bowman:Springer}.
All these actions have different systematic errors and the combination of 
these studies has given us an excellent handle on the possible lattice 
artifacts.  In every case, however, they have been performed in the quenched 
approximation and have been restricted to modest physical volumes.

In this paper we report first results for the quark propagator including
dynamical quark effects.  We use configurations generated by the MILC 
Collaboration~\cite{Bernard:2001av} available from the Gauge 
Connection~\cite{nersc}.  These use ``Asqtad'', \oa{2} improved staggered 
quarks~\cite{Orginos:1999cr}, giving us access to relatively light sea 
quarks.  In the quenched
approximation, the quark propagator for this action has excellent rotational
symmetry and is well behaved at large momenta~\cite{Bowman:2004xi}.  We use
quenched and dynamical configurations at the same lattice spacing and volume,
which enables us to observe the relatively subtle effects of unquenching. 
These lattices are also somewhat larger than those of previous studies, giving
us access to smaller momenta.

\section{Details of the calculation}

The quark propagator is gauge dependent and we work in the Landau gauge
for ease of comparison with other studies. 
Landau gauge is
a smooth gauge that preserves the Lorentz invariance of the theory, so
it is a popular choice.  It will be interesting to repeat this
calculation for the Gribov-copy free Laplacian gauge, and this is 
left for a future study.

The MILC configurations were generated with the \oa{2} one-loop
Symanzik-improved L\"{u}scher--Weisz gauge action~\cite{Luscher:1984xn}.
The dynamical
configurations use the Asqtad quark action, an \oa{2} Symanzik-improved
staggered fermion action.  
They have two degenerate light fermions, for the $u$ and $d$ quarks,
and a heavier one for the strange quark.
We explore a variety of light quark masses, 
with the bare strange quark mass  fixed at $ma = 0.05$, or $m$ = 79 MeV for 
$a$ = 0.125 fm~\cite{Davies:2003ik}.  In all cases the Asqtad 
action is also used for the valence quarks.  The values of the coupling and 
the bare sea-quark masses are matched such that the lattice spacing is held 
constant. This means that all systematics are fixed; the only variable is the 
addition of quark loops.  The simulation parameters are summarized in 
Table~\ref{simultab}.

\begin{table}[b!]
\caption{\label{simultab}Lattice parameters used in this study.
The dynamical configurations each have two degenerate light (up/down) quarks  
and a heavier (strange) quark. The lattice spacing is $a = 0.125(3)$ fm, 
where the uncertainty reflects
the variation of $a$ over the set of lattices considered in this
analysis. 
Bare light quark masses $ma = 0.01, 0.02, 0.03,0.04$
correspond to masses of $16 - 63$ MeV.  The bare strange quark 
mass is $ma = 0.05$ or $79$ MeV.}  
\begin{ruledtabular}
\begin{tabular}{ccccccccc}
   & Dimensions      & $\beta$  &Bare Quark Mass  & \# Config \\            
\hline
1  & $20^3\times 64$ &   8.00   &quenched              & 265 \\            
2  & $20^3\times 64$ &   6.76   &$16$ MeV, $79$ MeV  & 203 \\ 
3  & $20^3\times 64$ &   6.79   &$32$ MeV, $79$ MeV  & 249 \\           
4  & $20^3\times 64$ &   6.81   &$47$ MeV, $79$ MeV  & 268 \\       
5  & $20^3\times 64$ &   6.83   &$63$ MeV, $79$ MeV  & 318 
\end{tabular}
\end{ruledtabular}
\end{table}


On the lattice, the bare propagator $S(a; p^2)$ is related to the
renormalized propagator $ S^{\text{ren}}(\mu;p^2)$ through the
renormalization constant
\begin{equation}
S(a;p^2) = Z_2(a;\mu) S^{\text{ren}}(\mu;p^2) .
\end{equation}
In the continuum limit, Lorentz invariance allows one to decompose
the full quark propagator into Dirac vector and scalar pieces
\begin{equation}
S^{-1}(p^2) = i A(p^2) \gamma \cdot p + B(p^2)
\end{equation}
or, alternatively,
\begin{equation}
S^{-1}(p^2) = Z^{-1}(p^2) [i \gamma \cdot p + M(p^2)],
\end{equation}
where $M(p^2)$ and $Z(p^2)$ are the nonperturbative mass and wave function
renormalization functions, respectively.  Asymptotic freedom implies
that, as $p^2 \rightarrow \infty$, $S(p^2)$ reduces to the free
propagator
\begin{equation}
\label{eq:free_quark}
S^{-1}(p^2) \rightarrow  i\gamma \cdot p + m,
\end{equation}
up to logarithmic corrections. The mass function $M$ is renormalization point 
independent and for $Z$ we choose throughout this work the renormalization 
point as 3 GeV.

The tree-level quark propagator with the Asqtad action has the form
\begin{equation}
S^{-1}(p) = i \sum_\mu \gambar_\mu q(p_\mu) + m,
\end{equation}
where $q(p_\mu)$ is the kinematic momentum given by~\cite{Bowman:2002bm}
\begin{equation}
\label{eq:mom_Asqtad}
q_\mu \equiv \sin(p_\mu) \bigl[ 1 + \frac{1}{6} \sin^2(p_\mu) \bigr].
\end{equation}
The $\gambar_\mu$ form a staggered Dirac algebra.  Having
identified the kinematic momentum, we define the mass and
renormalization functions by
\begin{equation}
S^{-1}(p)  = Z^{-1}(q) \Bigl [ i \sum_\mu {(\gambar_\mu)}
     q_\mu(p_\mu) + M(q) \Bigr].
\end{equation}
Complete details of the extraction of the mass and renormalization
functions from the Asqtad propagator are described in the appendix.

\section{Quenched Results}

\begin{figure}[t]      
\centering\includegraphics[height=0.99\hsize,angle=90]{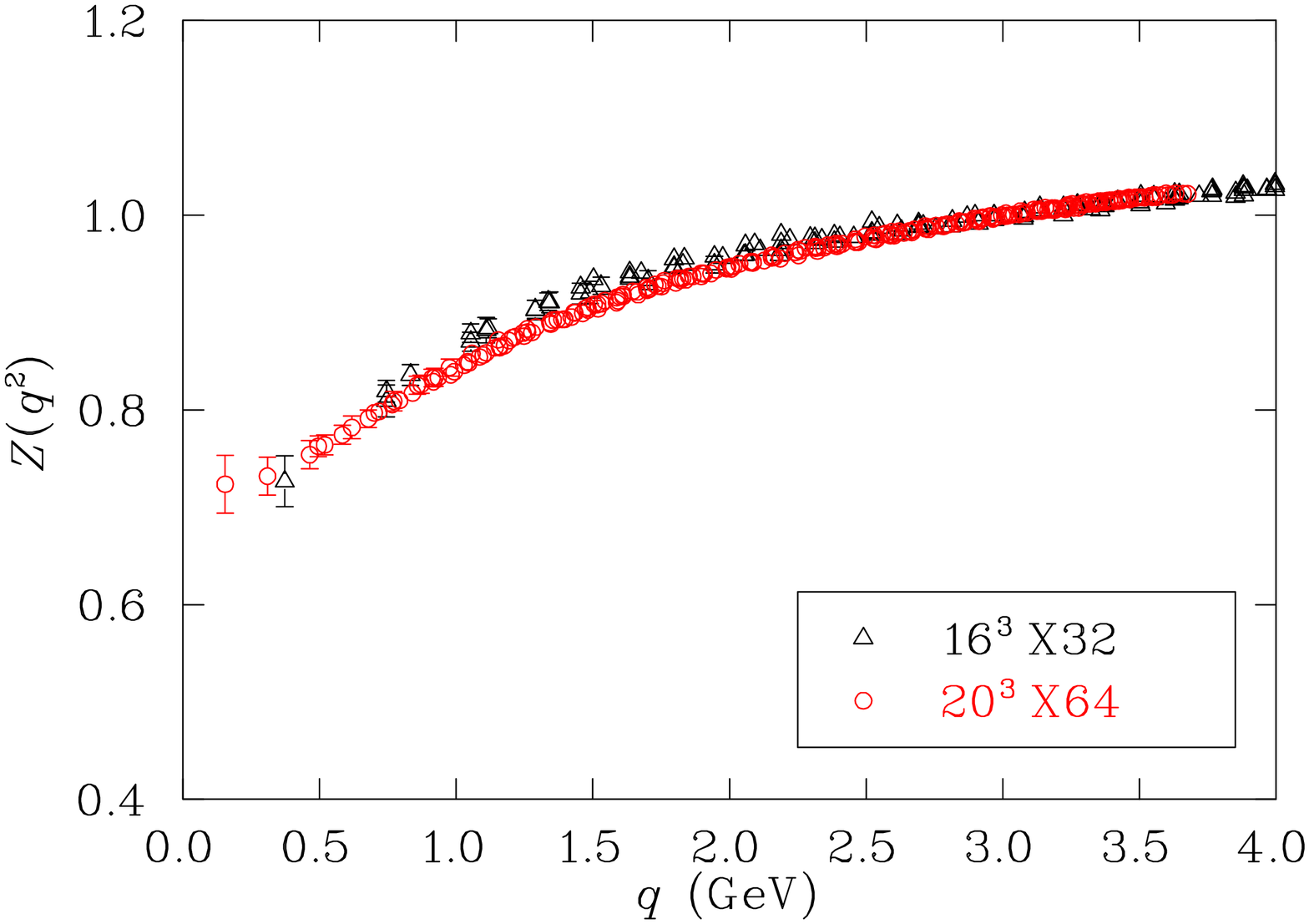}
\centering\includegraphics[height=0.99\hsize,angle=90]{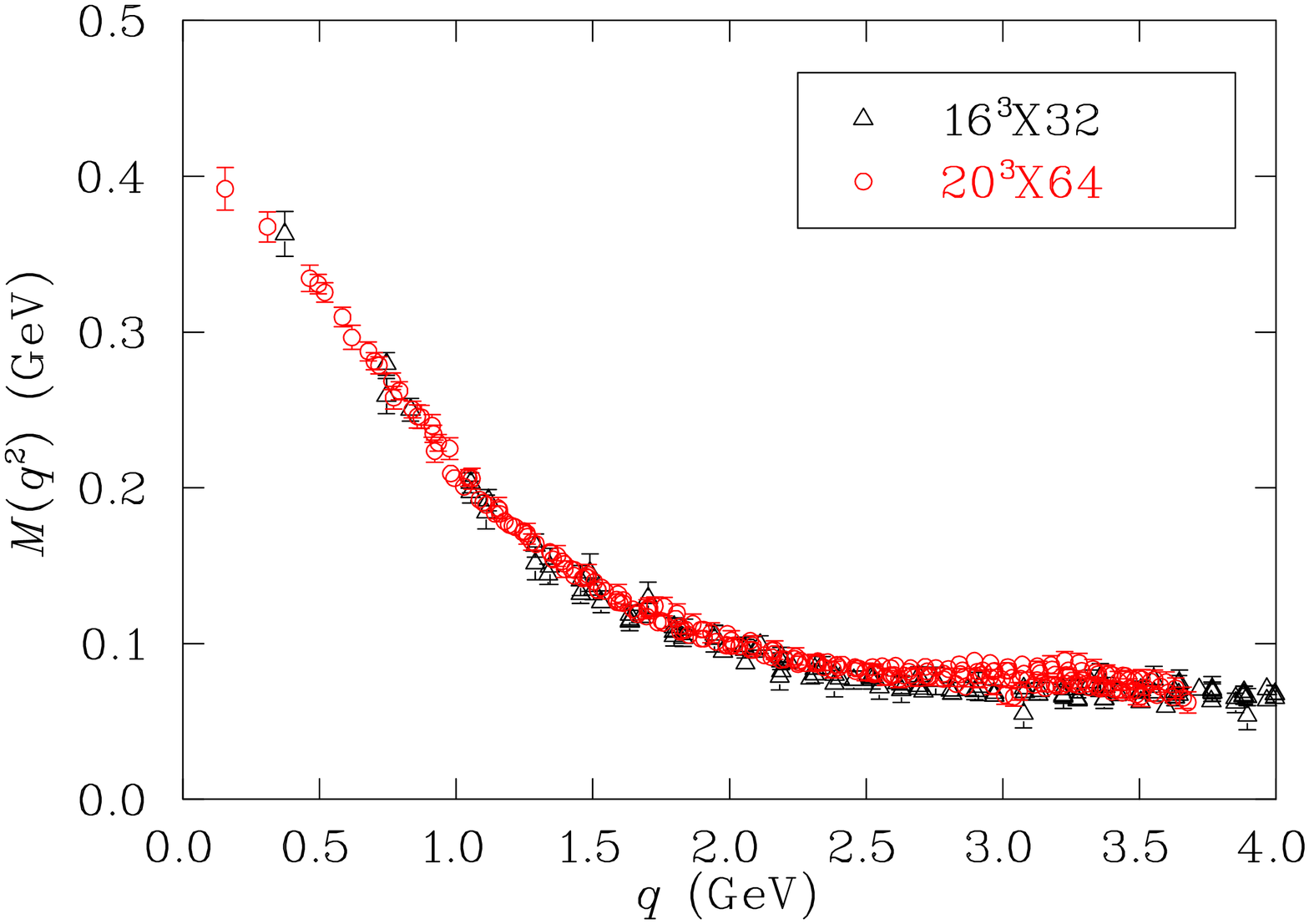}
\caption{Comparison of quenched wave-function renormalization and mass
functions at approximately the same bare quark mass.  The quark
propagator from the $20^3\times 64$ lattice with lattice spacing $a
=0.125$ fm at $m = 47$ MeV (open circles) is compared with the
previously published quark propagator from a $16^3\times 32$ lattice
with lattice spacing $a =0.105$ fm at $m = 45$ MeV (full triangles).
The renormalization point for $Z(q^2)$ is set at $q$ = 3 GeV.}
\label{quenched1}
\end{figure}

First we compare our quenched results to some previously published
data obtained on a smaller lattice~\cite{Bowman:2002kn}.  All the data
illustrated in the following are cylinder cut~\cite{Bonnet:2001uh}.
This removes points most susceptible to rotational symmetry breaking,
making the data easier to interpret.  As is well known, the definition
of lattice spacing in a quenched calculation is somewhat arbitrary,
and indeed the quoted estimate for our smaller ensemble is not
consistent with that published for the MILC configurations.  We
determined a consistent value of the lattice spacing by matching the
gluon propagator calculated on the old ensemble to that of the new
ensemble~\cite{Bowman:2004jm}.  This procedure yields a new nominal
lattice spacing of $a=0.105$ fm and physical volume of $1.7^3 \times
3.4~\text{fm}^4$ for the old lattices.  Examining the quark propagator
on the two quenched ensembles, shown in Fig.~\ref{quenched1}, we see
that the agreement is excellent.  This indicates that both finite
volume and discretization effects are small.  The flattening in the
deep infrared of both scalar functions is a long-standing prediction
of DSE studies~\cite{Bhagwat:2003vw}.

\begin{figure}[t]      
\centering\includegraphics[height=0.99\hsize,angle=90]{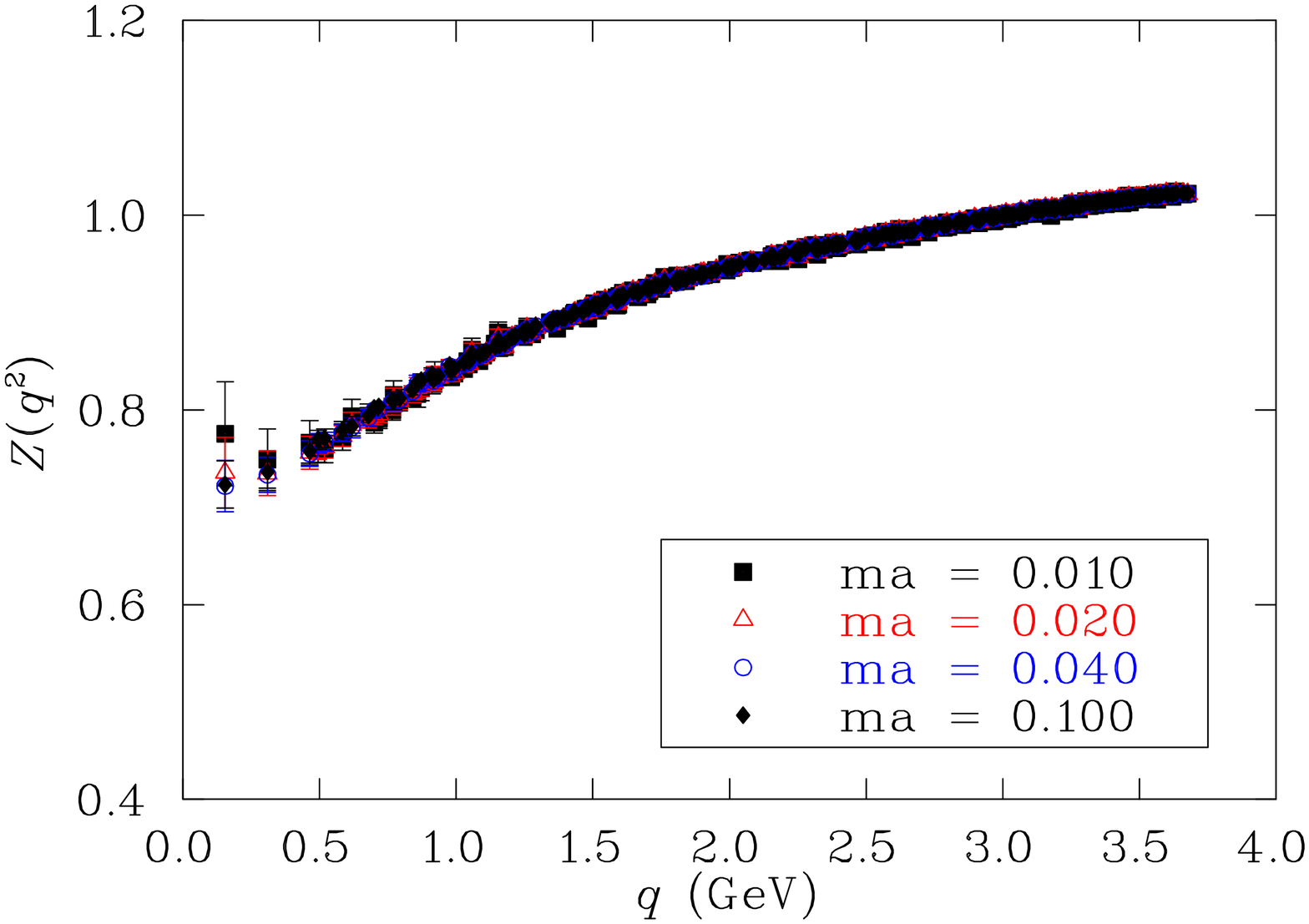}
\centering\includegraphics[height=0.99\hsize,angle=90]{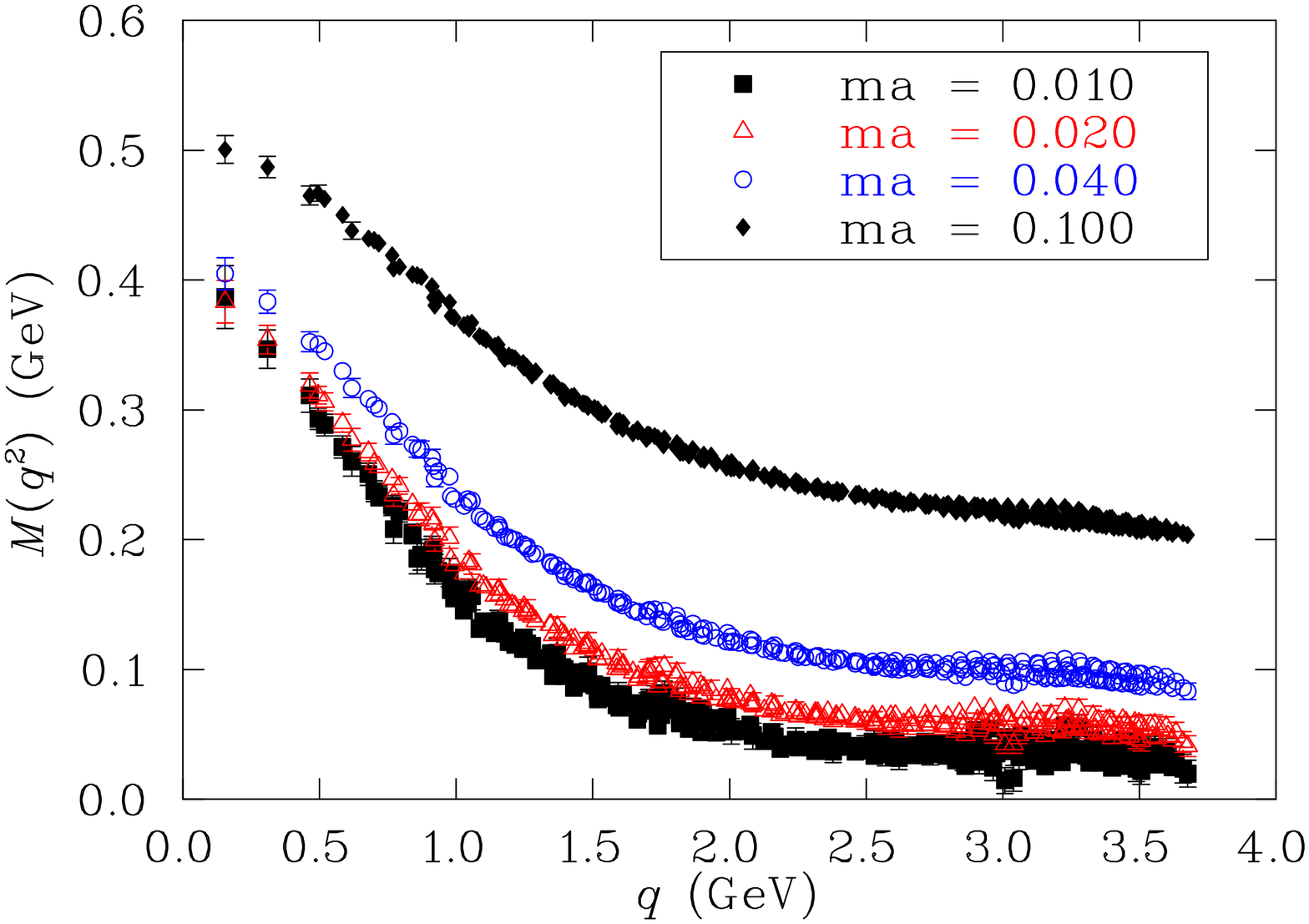}
\caption{The quenched renormalization function (top) and mass function
  (bottom) for a selection of quark masses, including $ma=0.100$,
  about twice the strange quark mass. The renormalization point for $Z(q^2)$ is set at $q$ = 3 GeV.}
\label{quenched2}
\end{figure}

We show results for the larger quenched lattice for a variety of bare
quark masses in Fig.~\ref{quenched2}. Once again we see that for quark
masses less than or approximately equal to that of the strange quark,
the lowest momentum point of the mass function is insensitive to quark
mass.

\section{Effects of Dynamical Quarks}

\begin{figure}[t]      
\centering\includegraphics[height=0.99\hsize,angle=90]{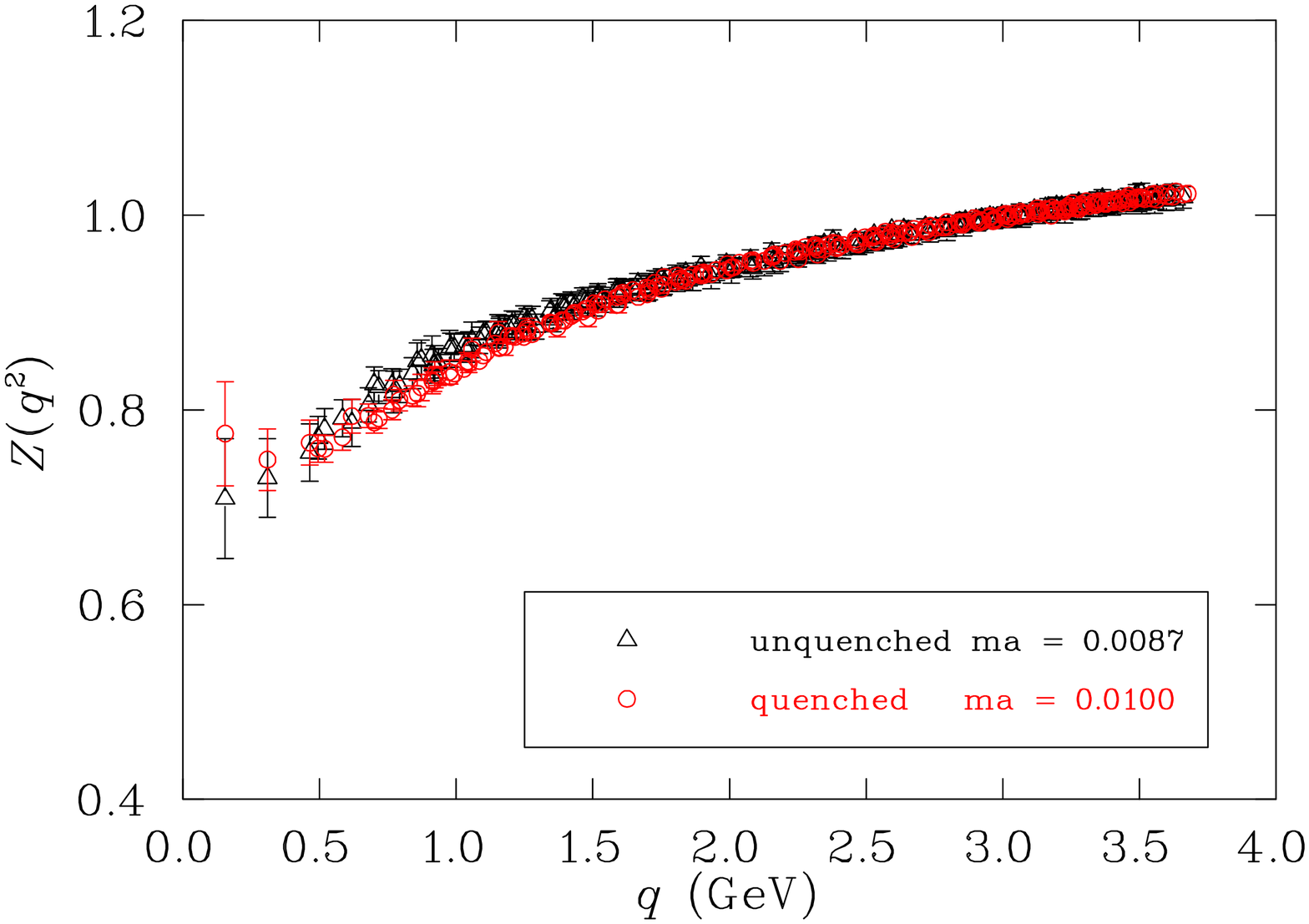}
\centering\includegraphics[height=0.99\hsize,angle=90]{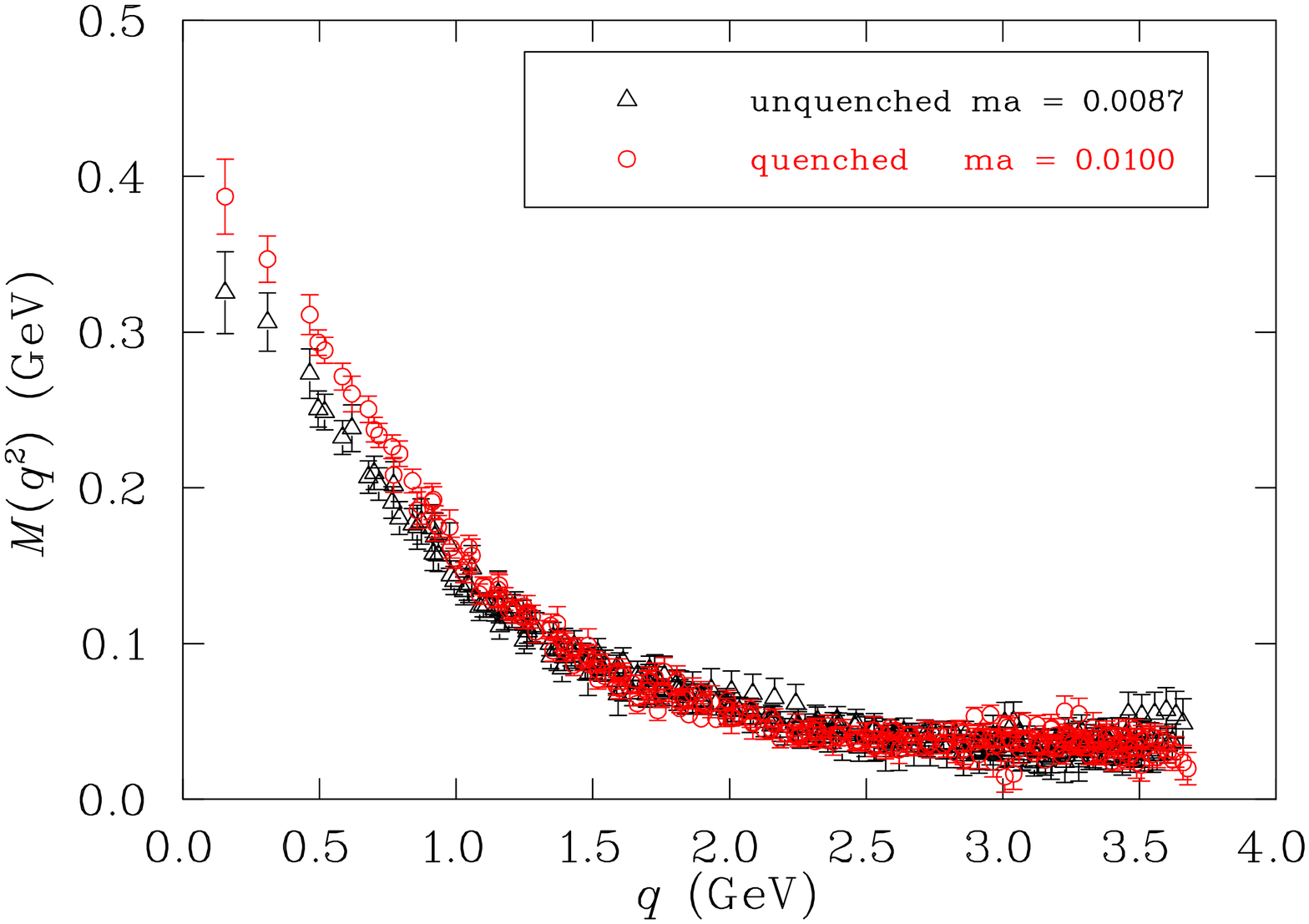}
\caption{Comparison of the unquenched (full QCD) and quenched quark
propagator for non-zero quark mass.  The mass function for the
unquenched dynamical-fermion propagator has been interpolated so that
it agrees with the quenched mass function for $ma = 0.01$ at the renormalization point, $q$ = 3 GeV.
For the unquenched propagator this corresponds to a bare quark mass of
$ma = 0.0087$.}
\label{withloops1}
\end{figure}

Here we compare the scalar functions for the quenched and dynamical
propagators.  For a given bare mass, the running mass depends upon
both the number of dynamical quark flavors and their masses.  To make
the most appropriate comparison we select a bare quark mass for the
quenched case ($ma = 0.01$) and interpolate the dynamical mass
function so that it agrees with the quenched result at the renormalization
point,  $q = 3$ GeV.  The results are shown
in Fig.~\ref{withloops1}.

The dynamical case does not differ greatly from the quenched case. For
the renormalization functions, there is no discernible difference
between the quenched and unquenched cases.  However the mass functions
do reveal the effects of dynamical quarks.  Dynamical mass generation
is suppressed in the presence of dynamical quarks relative to that
observed in the quenched case.  This is in accord with expectations as
the dynamical quark loops act to screen the strong interaction.  

\begin{figure}[t]
\centering\includegraphics[height=0.99\hsize,angle=90]{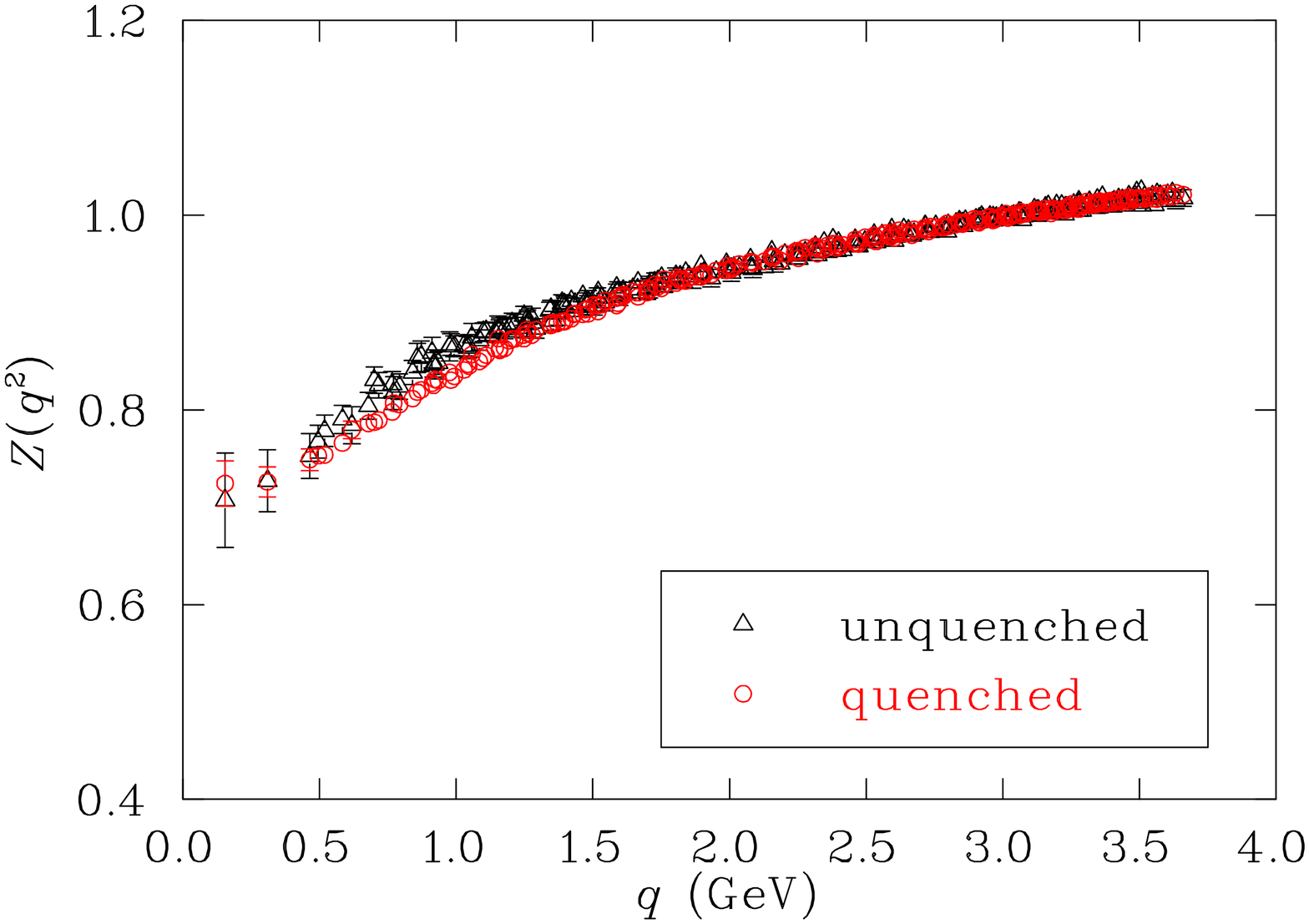}
\centering\includegraphics[height=0.99\hsize,angle=90]{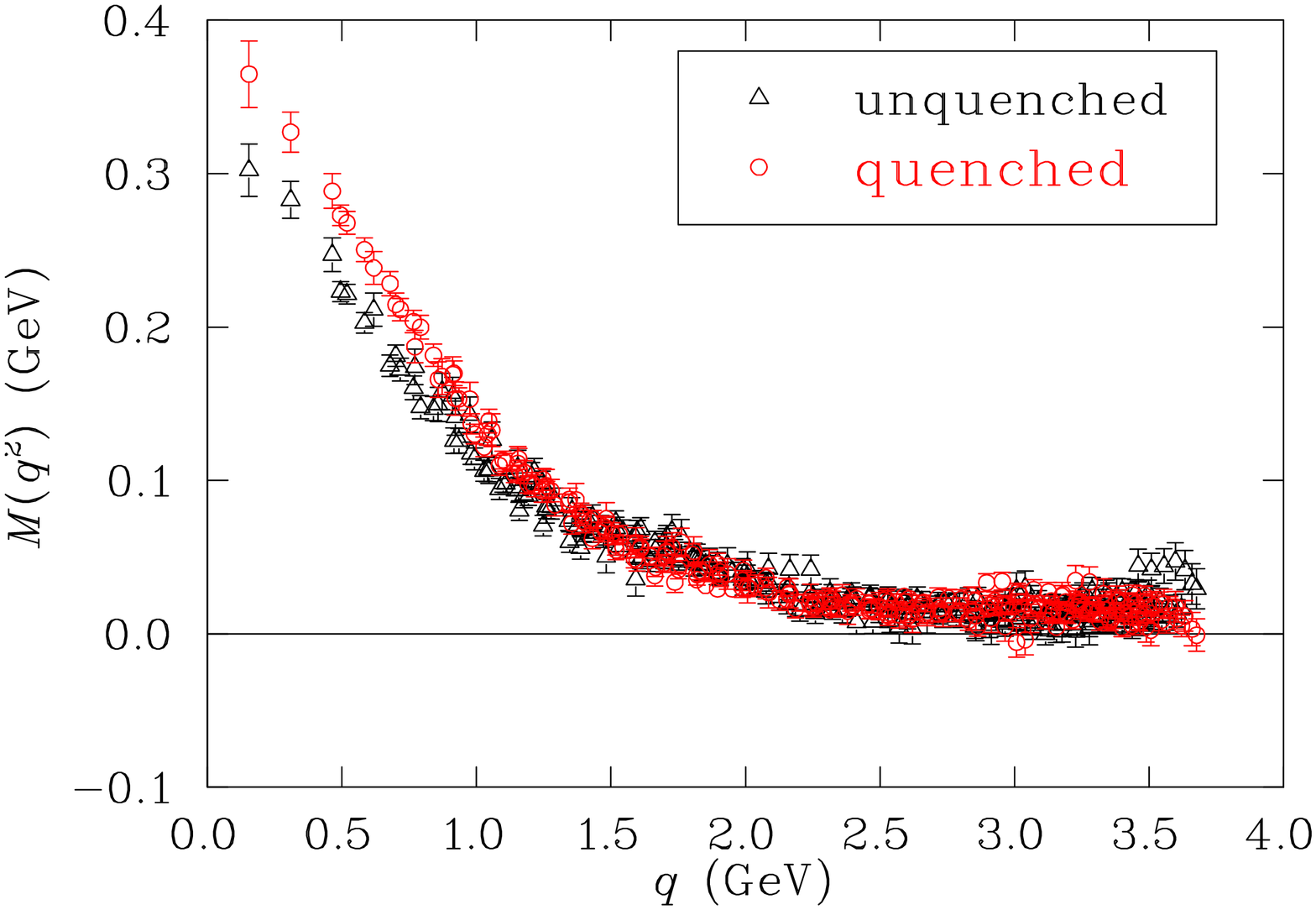}
\caption{Comparison of the unquenched (full QCD) and quenched quark
propagator in the chiral limit.  The renormalization function is
renormalized at $q$ = 3 GeV.  Whereas little difference is observed in the
renormalization function, the mass functions indicates that dynamical
mass generation is suppressed by the addition of quark loops.}
\label{withloops2}
\end{figure}

Further comparisons can be made in the chiral limit.  In
Fig.~\ref{withloops2} both quenched and dynamical data have been
extrapolated to zero bare quark mass by a fit linear in the quark
mass.  In the dynamical case, the extrapolation is limited to the case
where the valence and sea quark masses match for the light quarks.
While the results are qualitatively similar to those of
Fig.~\ref{withloops1}, increased separation is observed in the
generation of dynamical mass.  As discussed above, for a given bare
quark mass, the running mass is larger in full QCD than in quenched
QCD.

\begin{figure}[t]      
\centering\includegraphics[height=0.99\hsize,angle=90]{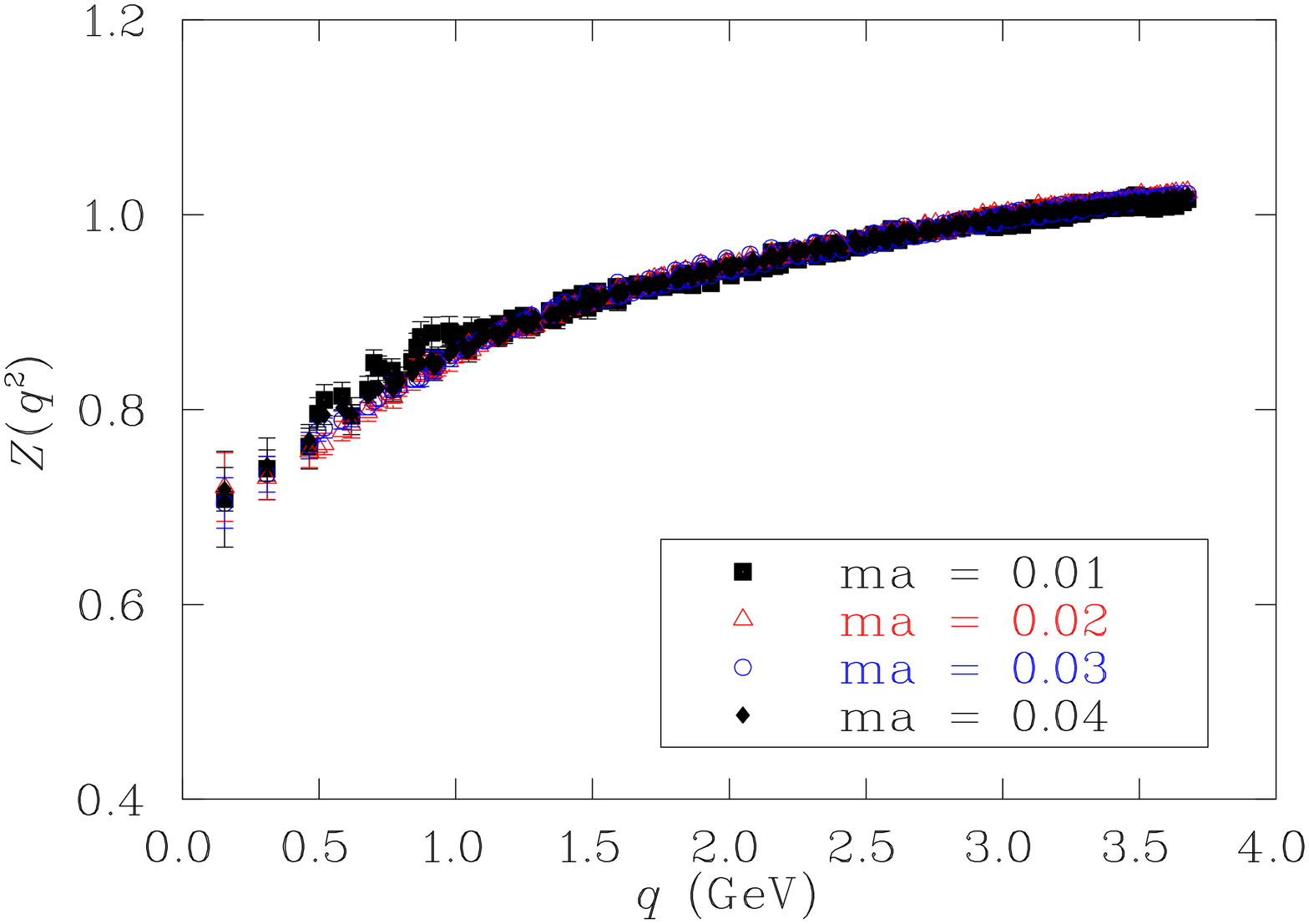}
\centering\includegraphics[height=0.99\hsize,angle=90]{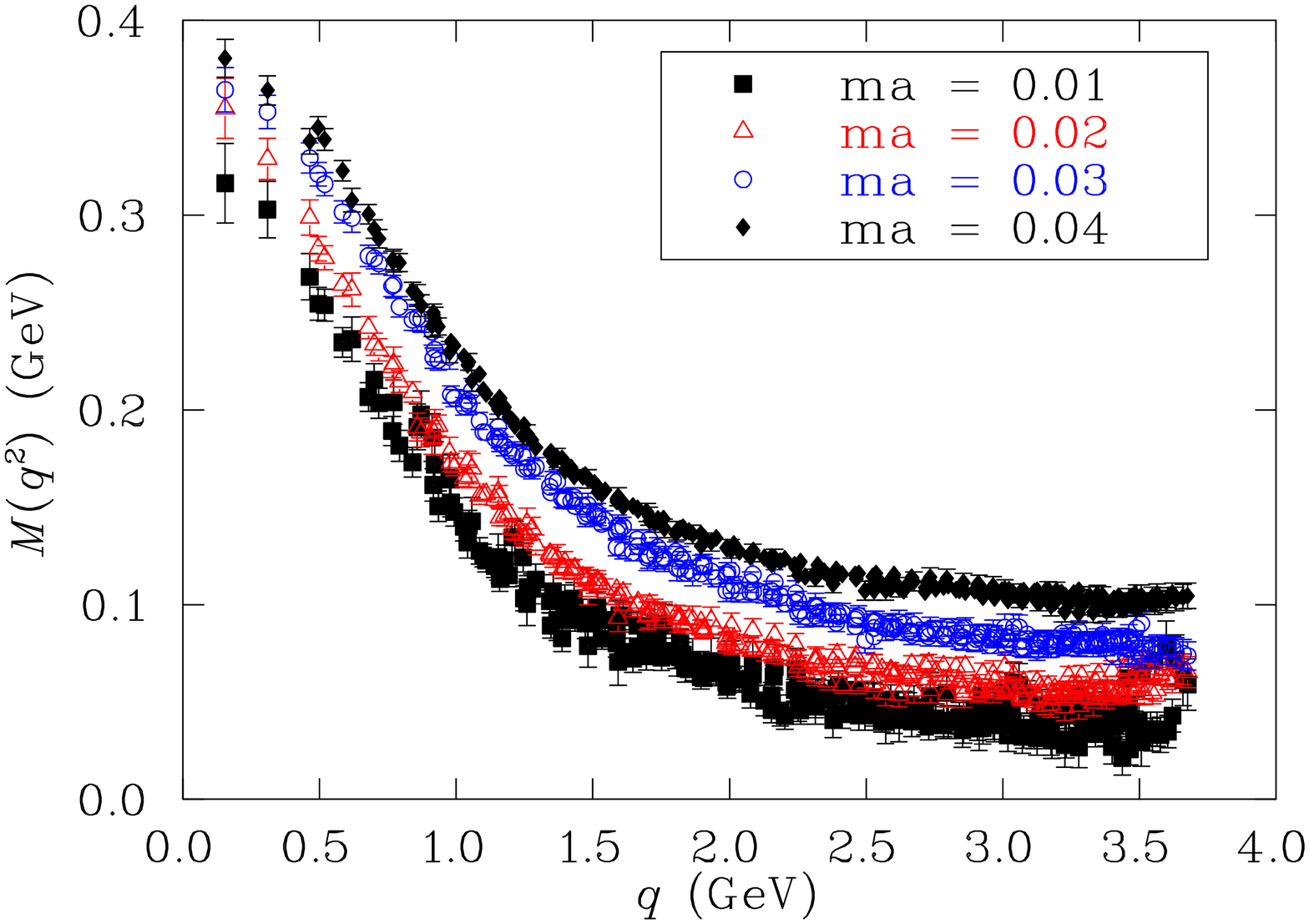}
\caption{Renormalization (top) and mass (bottom) functions for four
different quark masses in full QCD, where the valence and light-sea
masses are matched.  The wave-function renormalization function $Z$ is
renormalized at $q$ = 3 GeV.}
\label{alldyn}
\end{figure}

Fig.~\ref{alldyn} shows the mass and renormalization functions in the
dynamical case for a variety of quark masses.  Here the valence quark
masses and the light sea quark masses are matched.  The results show
that the renormalization function is insensitive to the bare quark
masses studied here.  The results for the mass function are ordered as
expected with the larger bare quark masses, $m$, giving rise to a
larger mass function.

%

\begin{figure}[t]      
\centering\includegraphics[height=0.99\hsize,angle=90]{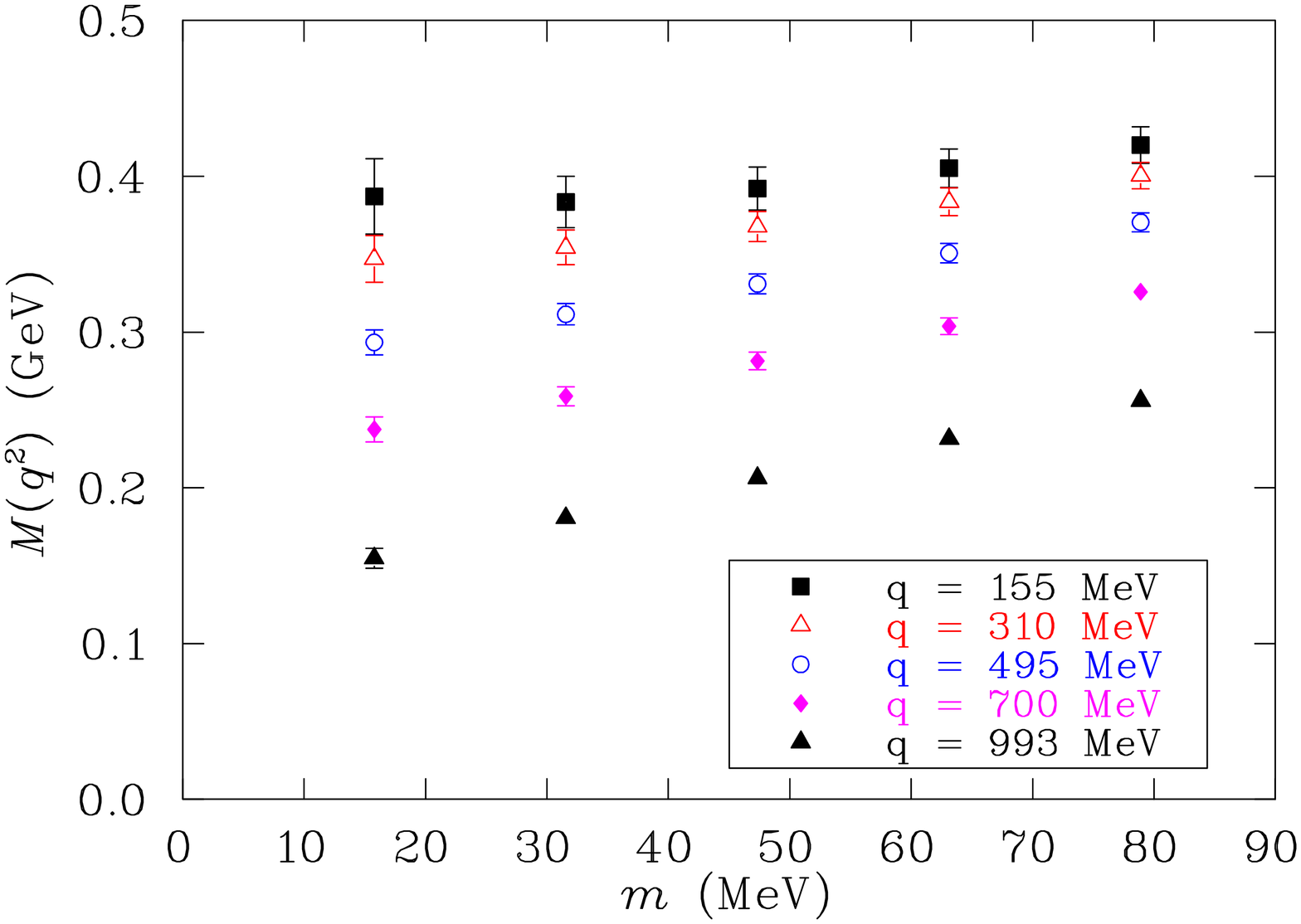}
\centering\includegraphics[height=0.99\hsize,angle=90]{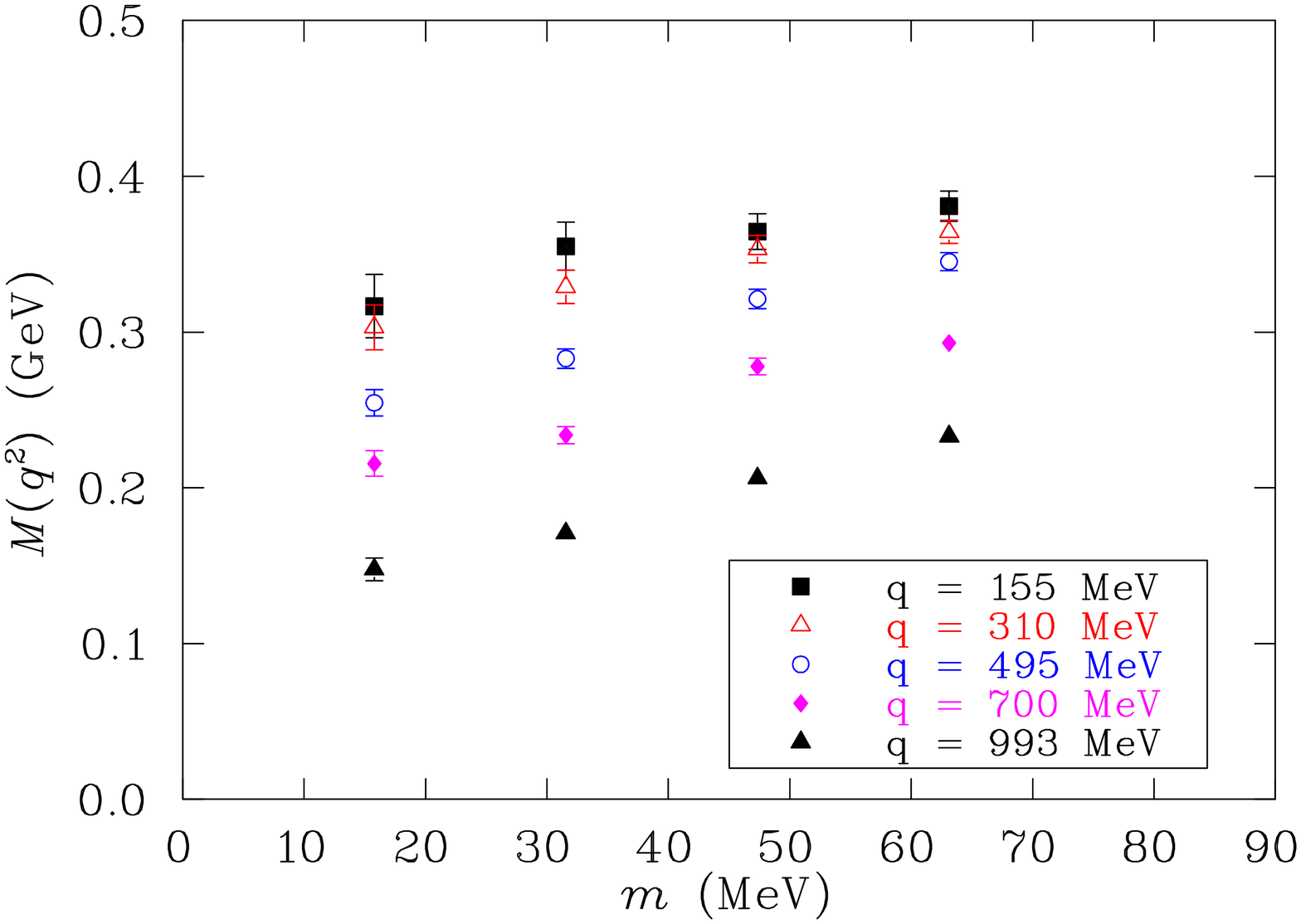}
\caption{The chiral limit approach of the mass function for selected
  momenta.  Results from quenched QCD simulations are illustrated at
  top whereas full dynamical-fermion QCD results are illustrated at
  bottom.  Non-linear behavior is observed for the lowest momentum
  points, in opposite directions for quenched and full QCD.  }
\label{extrap}
\end{figure}

Finally, we comment on the approach to the chiral limit.  In
Fig.~\ref{extrap} we show the mass function for five different momenta
plotted as a function of the bare quark mass.  The momenta considered
include the lowest momentum of 0.155 GeV and 0.310, 0.495, 0.700 and
0.993 GeV to explore momentum dependent changes in the approach to the
chiral limit.  At larger momenta, the mass function is observed to be
proportional to the bare quark mass.  However, at small momenta,
nonperturbative effects make this dependence more complicated.  For
example, a recent Dyson-Schwinger study predicts a downward turn as
the bare mass approaches zero~\cite{Bhagwat:2003vw}.

For the lowest momentum points, nonlinear behavior is indeed observed.
For the quenched case, curvature in an upward direction is revealed as
the chiral limit is approached, leading to the possibility of a larger
infrared mass function for the lightest quark mass, despite the
reduction of the input bare quark mass.  In contrast, a hint of downward
curvature is observed for the most infrared points of the full QCD
mass function as the chiral limit is approached.  It is interesting
that the nature of the curvature depends significantly on the 
chiral dynamics of the theory which are modified in making the
quenched approximation.  Similar behavior is observed in the hadron
mass spectrum where the coefficients of chiral nonanalytic behavior
can change sign in moving from quenched QCD to full QCD
\cite{Leinweber:2003ux,Young:2002cj}.

\section{Conclusions}

We have presented first results for the mass and wave-function
renormalization functions of the quark propagator in which the effects
of $2+1$ dynamical quark flavours are taken into account.  In contrast
to the significant screening suppression of the gluon propagator in
the infrared \cite{Bowman:2004jm}, the quark propagator is not
strongly altered by sea quark effects.  In particular, the
renormalization function is insensitive to the light bare quark masses
studied here, which range from 16 to 63 MeV, and also agrees well with
previous quenched simulation results.  Screening of dynamical mass
generation in the infrared mass function is observed when comparing
quenched and full QCD results at finite mass and in the chiral limit.
The approach of the mass function to the chiral limit displays
interesting nontrivial curvature for low momenta, with the curvature
in quenched and full QCD in opposite directions.

\section*{ACKNOWLEDGMENTS}

We thank the Australian National Computing Facility for Lattice Gauge
Theory and both the Australian Partnership for Advanced Computing (APAC)
and the South Australian Partnership for Advanced Computing (SAPAC)
for generous grants of supercomputer time which have enabled this
project.  This work is supported by the Australian Research Council.

\appendix*
\section{Extraction of the scalar functions}

The Asqtad quark action~\cite{Orginos:1999cr} is a staggered action
using three-link, five-link and seven-link staples as a kind of
``fattening'' to minimize quark flavor (often referred to as
``taste'') changing interactions.  The three-link Naik
term~\cite{Naik:1986bn} is included to improve rotational symmetry by
improving the finite difference operator, and the five-link Lepage
term~\cite{Lepage:1998vj} is included to correct errors at low momenta
that may be introduced by the above mentioned staples .  The
coefficients are tadpole improved and chosen to remove all tree-level
\oa{2} errors.

At tree-level (i.e. no interactions, links set to the identity) the staples
in this action make no contribution, so the action reduces to the tree-level
Naik action,
\begin{multline}
S^{(0)} = \frac{1}{2} \sum_{x,\mu} \chibar(x) \eta_\mu(x) \Bigl[
   \frac{9}{8} \bigl( \chi(x+\mu) - \chi(x-\mu) \bigr) \\ - 
   \frac{1}{24} \bigl( \chi(x+3\mu) - \chi(x-3\mu) \bigr) \Bigr]
   + m\sum_x \chibar(x)\chi(x),
\end{multline}
where the staggered phases are: $\eta_\mu(x) = (-1)^{\zeta^{(\mu)} \cdot x}$ 
and
\begin{equation}
\zeta^{(\mu)}_\nu = \biggl\{ \begin{array}{ll}
			1 & \mbox{if $\nu < \mu$} \\
			0 & \mbox{otherwise.}
		 \end{array}
\end{equation}
In momentum space, the quark propagator with this action has the tree-level 
form
\begin{multline}
\label{eq:Asq_tree}
{S^{(0)}_{\alpha\beta}}^{-1}(p;m)  \\
= i \sum_\mu (\gambar_\mu)_{\alpha\beta}
    \Bigl[\frac{9}{8} \sin(p_\mu) 
       - \frac{1}{24} \sin(3p_\mu) \Bigr] + m\deltabar_{\alpha\beta}~~~~~~ \\
 = i \sum_\mu (\gambar_\mu)_{\alpha\beta} \sin(p_\mu) \bigl[ 1  
       + \frac{1}{6} \sin^2(p_\mu) \bigr]  + m\deltabar_{\alpha\beta} .
\end{multline}
where the $\alpha,\beta$ are themselves four-vectors: $\alpha_\mu =
0,1$, and likewise for $\beta$; thus the quark propagator in
\eref{eq:Asq_tree} is a $16\times16$ matrix.  This familiar form is
obtained by defining
\begin{align}
\label{eq:deltabar}
\deltabar_{\alpha\beta} & = \Pi_\mu \delta_{\alpha_\mu\beta_\mu |\bmod 2} \\
\label{eq:gambar}
{(\gambar_\mu)}_{\alpha\beta} & = (-1)^{\alpha_\mu} 
	\deltabar_{\alpha+\zeta^{(\mu)},\beta}.
\end{align}
The $\mod~2$ in \eref{eq:deltabar} ensures its validity in \eref{eq:gambar}.
The $\gambar_\mu$ satisfy
\begin{gather}
\{\gambar_\mu, \gambar_\nu\}_{\alpha\beta} = 2 \delta_{\mu\nu} 
     \deltabar_{\alpha\beta} \\
\gambar_\mu^\adj = \gambar_\mu^T = \gambar_\mu^* = \gambar_\mu,
\end{gather}
forming a ``staggered'' Dirac algebra.  

Staggered actions are invariant under translations of 
$2a$, and the momentum on this blocked lattice is
\begin{equation}
\label{eq:stag_mom}
p_\mu = \frac{2\pi m_\mu}{L_\mu} \quad | 
   \quad m_\mu = 0,\ldots,\frac{L_\mu}{2}-1.
\end{equation}
We calculate the quark propagator in coordinate space,
\begin{equation}
G(x,y) = \langle \chi(x)\chibar(y) \rangle,
\end{equation}
and obtain the quark propagator in momentum space by Fourier transform of 
$G(x,0)$.  To write the Fourier transform of the staggered field we 
write the momentum on the lattice
\begin{equation}
k_\mu = \frac{2\pi n_\mu}{L_\mu} \quad | \quad n_\mu = 0,\ldots,L_\mu-1
\end{equation}
so that $k_\mu = p_\mu + \pi\alpha_\mu$ and define 
$\int_k \equiv \frac{1}{V} \sum_k$.  Then
\begin{gather}
\int_k = \int_p \sum_{\alpha_\mu = 0}^1, \\
\chi(x) = \int_k e^{ik \cdot x}\chi(k) = \int_p \sum_{\alpha} 
   e^{i(p+\pi\alpha) \cdot x} \chi_\alpha(p),
\end{gather}
and
\begin{align}
G(x,y) &= \sum_{\alpha\beta} \int_{p,l} \exp{\bigl\{ i(p+\alpha\pi)\cdot x 
  - i(l+\beta\pi)\cdot y\bigr\}} \nonumber\\
  &\qquad \times \langle \chi_\alpha(p)\chibar_\beta(l) \rangle \\
       &=\sum_{\alpha\beta} \int_{p} e^{ip(x-y)} 
         e^{i\pi(\alpha\cdot x - \beta\cdot y)} S_{\alpha\beta}(p). 
\end{align}
%
Now it will be convenient to re-write this
\begin{align}
G(k) & = G(l+\pi\delta) \equiv G_\delta(l) = \sum_x e^{-ikx} G(x,0) \notag\\
	&= \sum_{\alpha\beta} \int_p \sum_x \exp\bigl\{ -i(l+\pi\delta)x 
	\bigr\} \notag\\
	  &\qquad\qquad \times \exp\bigl\{ i(p+\pi\alpha)x \bigr\} 
	  S_{\alpha\beta}(p) \notag\\
\label{eq:mom_prop}
	&= \sum_{\alpha\beta} \int_p \delta_{pl}\deltabar_{\alpha\delta}
	S_{\alpha\beta}(p) \\
	&= \sum_\beta S_{\delta,\beta}(l).
\end{align}

In the interacting case, the quark propagator asymptotically approaches its 
tree-level value due to asymptotic freedom.  At finite lattice spacing the
actual behavior is closer to 
\begin{equation}
S(q;m) \rightarrow \frac{1}{u_0} S^{(0)}(q; m/u_0)
\end{equation}
where $u_0$ is the tadpole (or mean-field) improvement factor defined by
\begin{equation}
u_0 = (<\tr\, U_{\text{plaq}} >)^{1/4} .
\end{equation}

Assuming that the full lattice propagator retains its free form (in analogy 
to the continuum case) we write
\begin{align}
S_{\alpha\beta}^{-1}(p) &= i\sum_\mu {(\gambar_\mu)}_{\alpha\beta} 
   q_\mu(p_\mu) A(p) + B(p) \deltabar_{\alpha\beta} \\
	&= Z^{-1}(p) \Bigl[i\sum_\mu {(\gambar_\mu)}_{\alpha\beta} 
     q_\mu(p_\mu) + M(p) \deltabar_{\alpha\beta} \Bigr],
\end{align}
where $q$ is the tree-level momentum, \eref{eq:mom_Asqtad}.  Combining this 
with \eref{eq:mom_prop} above, we can extract the scalar functions (which we 
now write in terms of $q$) as follows:
\begin{equation}
G_\alpha(q) = Z(q) \frac{-i\sum_\mu (-1)^{\alpha_\mu} q_\mu + M(q)}{q^2 + M^2(q)},
\end{equation}
from which we obtain
\begin{align}
\sum_\alpha \tr\, G_\alpha(q)  
	&= 16N_c \frac{Z(q)M(q)}{q^2 + M^2(q)} \notag\\
	&= 16N_c {\cal B}(q),
\end{align}
and
\begin{align}
i\sum_\alpha\sum_\mu (-1)^{\alpha_\mu} q_\mu \tr\, [G_\alpha(q)]
	&= 16N_c q^2 \frac{Z(q)}{q^2 + M^2(q)} \notag\\
	&= 16N_c q^2 {\cal A}(q).
\end{align}

Putting it all together we get
\begin{align}
A(q) &= Z^{-1}(q) = \frac{{\cal A}(q)}{{\cal A}^2(q) q^2
	+ {\cal B}^2(q_\mu)} \\
B(q) &= \frac{M(q)}{Z(q)} = \frac{{\cal B}(p)}
	{{\cal A}^2(q) q^2 + {\cal B}^2(p)} \\
M(q) &= \frac{{\cal B}(q)}{{\cal A}(q)}.
\end{align}
By calculating ${\cal A}, {\cal B}$ instead of $A, B$, we avoid inverting the
propagator.  We calculate the ensemble average of ${\cal A}$ and ${\cal B}$ 
and thence $M$ and $Z$.


\end{document}